\pdfoutput=1
\documentclass[journal]{IEEEtran}

\usepackage{cite}
\usepackage{amsmath,amssymb,bm,bbm}
\usepackage{xcolor}
\usepackage[mathscr]{euscript}
\usepackage[pdftex]{graphicx}
\usepackage{algorithm}
\usepackage{epstopdf}
\usepackage{layouts}

\usepackage{algorithmic}

\newcommand{\norm}[1]{\left\Vert{#1}\right\Vert}
\newcommand{\CRB}{Cram\'{e}r Rao Bound }
\newcommand{\ZZB}{Ziv Zakai Bound }
\newcommand{\Ebb}{\mathbb{E}}

\begin{document}
\title{Design of Large Effective Apertures for Millimeter Wave Systems using a Sparse Array of Subarrays}
%
\author{Anant~Gupta,~\IEEEmembership{Student Member,~IEEE},
        Upamanyu~Madhow,~\IEEEmembership{Fellow,~IEEE}
        Amin Arbabian,~\IEEEmembership{Senior Member,~IEEE}
        and~Ali Sadri
\thanks{A. Gupta and U. Madhow are with the Department
of Electrical and Computer Engineering, University of California Santa Barbara, CA USA (Email: \{anantgupta, madhow\}@ece.ucsb.edu)} 
\thanks{A. Arbabian is with the Department
of Electrical Engineering, Stanford University, CA USA (Email: arbabian@stanford.edu)}
\thanks{A. Sadri is with Intel Corporation (Email: ali.s.sadri@intel.com}}

\markboth{IEEE Transactions on Signal Processing}
{Shell \MakeLowercase{\textit{Gupta et al.}}: Aperture Extension using Subarrays}
%



\maketitle

\begin{abstract}
We investigate synthesis of a large effective aperture using a sparse array of subarrays. We employ a multi-objective optimization framework for placement of subarrays within a prescribed area dictated by form factor constraints, trading off the smaller beam width obtained by spacing out the subarrays against the grating and side lobes created by sparse placement. We assess the performance of our designs for the fundamental problem of bearing estimation for one or more sources, comparing performance against estimation-theoretic bounds.  Our tiled architecture is motivated by recent progress in low-cost hardware realizations of moderately sized antenna arrays (which play the role of subarrays) in the millimeter wave band, and our numerical examples are based on 16-element ($4 \times 4$) subarrays in the 60 GHz unlicensed band.

\end{abstract}


%
\IEEEpeerreviewmaketitle

\section{Introduction} \label{sec:intro}

Many sensing and situational awareness applications (e.g., radar imaging for vehicles and drones) require highly directional, electronically steerable beams.  Reducing beam width requires expansion of antenna aperture. This is typically accomplished by filling the aperture with antenna elements spaced at half the carrier wavelength or less, in order to avoid grating lobes.  However, this approach does not scale well with aperture size since the cost, power consumption and design complexity increases with number of antenna elements.  

In this paper, we investigate the problem of synthesizing narrow beams using a tiled architecture, with a sparse set of subarrays spread over a large physical aperture.  Each subarray is a relatively compact antenna array with a moderate number of elements. The total
number of antenna elements, summing over subarrays,  is much smaller than that for a classical design spanning the entire physical aperture.  The sidelobes and grating lobes resulting from such spatial undersampling must therefore be controlled in order for our proposed ``array of subarrays'' to be useful.  Our goal here is to determine the placement of a given number of subarrays over a physical aperture in order to optimize multiple beam attributes, including beam width, maximum sidelobe level, and directivity.  

While our framework is general, the design of millimeter wave arrays is of particular interest to us, because the small carrier wavelength enables synthesis of narrow beams using relatively compact apertures. As a running example throughout this paper, we consider the design of a 60 GHz array of subarrays created by placing 8 subarrays over an aperture size of $10$ cm by $10$ cm ($20\lambda \times 20\lambda$ for $\lambda=0.5$ cm), where each subarray has $4\times 4$ elements arranged in uniform rectangular grid with $0.5\lambda$ horizontal spacing and $0.6\lambda$ vertical spacing. The subarrays need to be aligned along their axes, assuming that all elements have unidirectional linear polarization.  Each subarray tile occupies extra physical area on the plane, which must also be accounted for in the placement procedure. These design restrictions are consistent with existing prototype subarrays. 

\subsection{Contributions} 

Our contributions are summarized as follows:\\
$\bullet$ We formulate the problem of subarray placement as multi-objective optimization of key performance measures such as beam width, maximum sidelobe level, eccentricity and directivity,
\begin{equation}
\label{objectives}
\begin{aligned}
\text{Minimize }&\quad \mathtt{BW}(\bm{C,w}),\mathtt{MSLL}(\bm{C,w}), \mathtt{ecc}(\bm{C,w})\\
\text{Maximize }& \quad G_D(\bm{C,w})\\
\text{subject to}& \quad \textsc{AoS}(\bm{C})
\end{aligned}
\end{equation}
where $\bm{C}$ is $N_{s}\times2$ Subarray center position matrix, $\bm{w}$ is $N\times1$ beamsteering weight vector and $\textsc{AoS}(\bm{C})$ are physical  constraints to avoid overlapping subarrays.  Note that orientation of subarrays is not an optimization variable in this architecture, since the polarization of the elements has to be aligned for beamforming.
The problem (\ref{objectives}) is non-convex and combinatorially explosive. We use geometric heuristics to eliminate similar configurations in the first stage of our algorithm, and then employ a second stage of refinement using small perturbations around the first stage solution. \\
$\bullet$ We evaluate our designs using estimation-theoretic benchmarks for two-dimensional (2D) direction of arrival (DoA) estimation.
At low signal-to-noise ratio (SNR), large sidelobes can lead to large errors in the DoA estimate.  At high SNR, on the other hand, the DoA estimation error is governed by beam width.  We derive a Ziv-Zakai bound (ZZB), which captures the effect of both
large and small estimation errors, for DoA estimation for specular paths.  The ZZB exhibits a distinct transition in its behavior from low to high SNR, tending
at high SNR to the Cramer-Rao bound (CRB),
which captures the effect of small errors around the true parameter value.  Thus, we use the ZZB transition SNR as a measure of efficacy of
sidelobe reduction, and the CRB as a measure of efficacy of beam width reduction.\\
$\bullet$ The beam attributes of arrays designed using multi-objective optimization depend on its parameterization.  We design two sparse arrays, $A1$ with primary emphasis on reducing beamwidth and $A2$ based on joint optimization of beamwidth and maximum sidelobe level. These designs are compared against two benchmark arrays. The first is termed a ``compact array,'' with subarrays packed closely together: this is expected to have worse beam width but smaller sidelobes than our sparse designs.  
The second is termed a ``naive array,'' obtained by placing subarrays in diamond pattern to obtain beamwidth equivalent to that of sparse array \emph{A2}.  Some illustrative numerical results for our running example are as follows. The sparse design \emph{A1} is $11$ dB better than the compact array in terms of CRB,
while degrading less than $1$ dB in terms of ZZB threshold. The sparse design \emph{A2} is $4$ dB better in terms of CRB than the ``naive array,'' while also having a better ZZB threshold.\\ 
$\bullet$ We investigate DoA estimation performance numerically using a state of the art super-resolution algorithm.  The impact of the higher sidelobes due to sparse placement, and hence that of our optimization procedure, is more evident when estimating DoA in the presence of multiple interfering targets. 
We show that, depending on the strength of interferers, our optimized arrays achieve better estimation accuracy than the ``compact'' and ``naive'' benchmark arrays at moderate to high SNR, due to a combination of sharper beamwidth and lower sidelobes. We also show that the efficacy of DoA estimation using our sparse designs, and the associated benchmarks, is maintained when we employ compressive measurements.

\subsection{Related Work}

There is a rich body of work on sparsifying linear arrays, including minimum redundancy arrays \cite{moffet1968minimum}, genetic optimization\cite{birinci2005optimization}, joint \CRB and sidelobe level optimization \cite{roy2013sparsity}, and simulated annealing  \cite{trucco1999thinning}. Most popular design strategies try to find an element pattern which minimizes beamwidth, along with some notion of DoA ambiguities such as sidelobe level or probability of DoA outlier.  Recent approaches like Nested 2D arrays \cite{coprime_array} and H-arrays \cite{keto2012hierarchical} utilize the idea of ``difference co-array'' to reduce the number of redundant spacings and maximize the randomness of element positions, so that the number of spatial frequencies being sampled by the array is maximized.  Most of these techniques, however, assume that antenna elements can be placed freely.  Hence, they do not apply in our setting,
where element placement within subarrays is constrained. 

The prior work most similar to our is \cite{athley2007radar}, which investigates design of linear arrays with two and three subarrays. However, the focus there is on performance criteria for comparing a number of sensible designs in a far smaller design space, rather than searching over a large space of possibilities as we do here.

Our performance evaluation requires implementation of DoA estimation algorithms.  Classical subspace-based algorithms such as MUSIC \cite{music} and ESPRIT \cite{esprit}, as well as their
extensions to arrays of subarrays such as  \cite{wong1998direction,zoltowski2000closed,haardt1998simultaneous,vasylyshyn2005direction}, rely on regular array geometries for efficient computation. Recently developed super-resolution algorithms such as Basis Pursuit Denoising (BPDN) \cite{van2011sparse} 
and Newtonized Orthogonal Matching Pursuit (NOMP) \cite{mamandipoor2016newtonized} are both more general and have better performance. It is worth noting that \cite{stoeckle2015doa} shows that, for large arrays, BPDN and other sparse estimation techniques with compressive measurements outperform subspace-based methods. In this paper, we employ NOMP in our numerical experiments, since we have found it to 
provide better performance than BPDN at lower complexity. We also show that the performance trends are unchanged under compressive measurements, consistent with recent general theory \cite{ramasamy2014compressive}.

\subsection{Outline} 

We first describe the beam attributes to be optimized while designing the sparse arrays and discuss constraints for the optimization in Section \ref{sec:sparse_design}.  The geometric heuristics and design approach is described in detail in Section \ref{placement_opt}. We then provide a brief review of estimation bounds for 2D bearing estimation and discuss their utility for analysing the sparse arrays in Section \ref{estimation}. 
Finally, we compare the example array configurations against benchmarks and evaluate the performance of super-resolution algorithms in Section \ref{results_all}.

\section{Sparse Subarray Design} \label{sec:sparse_design}

We formulate the array design problem in terms of jointly optimizing multiple beam parameters that are expected to affect DoA estimation performance.

\begin{figure}[htbp]
\centering
\includegraphics[width=\columnwidth]{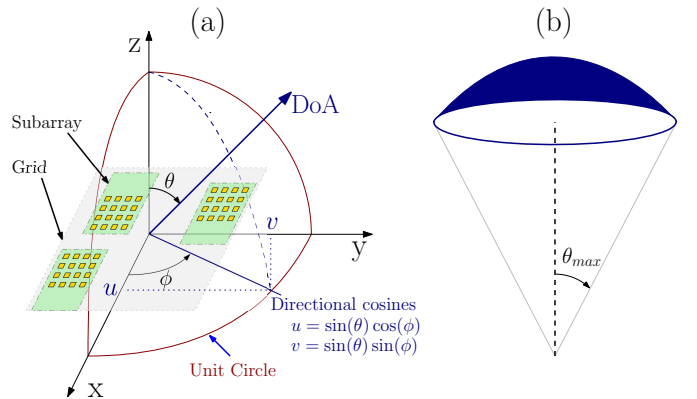}
\caption{(a) 2D Array Geometry and Spherical coordinate system. (b)ROI: Uniform distribution of 2D-DoA $\bm{u}$ in spherical cap with half angle $\theta_{max}$}
\label{DoA}
\end{figure}

\subsection{Beam Pattern Basics}
\label{bp_basics}
We use the directional cosines 
\begin{equation} \label{uv_defn}
u=\sin(\theta)\cos(\phi), v=\sin(\theta)\sin(\phi)
\end{equation}
to represent the DoA of target. The elevation $\theta$ and azimuth $\phi$ angles are measured from the broadside direction (perpendicular to the baseline array plane). The 2D beampattern $R(u,v)$ in direction $(u,v)$ when the beam is steered towards the broadside is given by
\begin{equation}
\label{beampattern_eq}
R(u,v)=\frac{1}{N^2}\left | \sum_{i=1}^N {e^{jk\left( u d_i^x+ v d_i^y \right)}}\right |^2
\end{equation}
where $N$ is the number of array elements; $[d_i^x,d_i^y]^T\triangleq \bm{d}_i$ are the 2D co-ordinates of arrays elements, and $k=\frac{2\pi}{\lambda}$ is the wavenumber. We assume isotropic antenna elements with ideal steering weights and far-field sources with normalized response.  The term ``subarray'' refers to the subset of elements with uniform half-wavelength spacing, while ``super-array'' refers to the placement of these subarrays, which is described by the subarray centers.  Since the elements in a subarray are fixed, the array element locations, $\bm{D}$ can be expressed in terms of the subarray centers, $\bm{C}$, as $\bm{D}=\bm{C}\otimes \mathbbm{1}_{N_e} + \bm{D}_e\otimes \mathbbm{1}_{N_{s}}$, where $\bm{D}_e$ is the fixed $2\times N_e$ matrix containing the subarray element coordinates with respect to its center, $N_{s}$ is the number of subarrays, $N_{e}$ is number of elements in individual subarrays, $\mathbbm{1}_n$ is an $n\times 1$ column vector of ones, and $\otimes$ denotes the Kronecker product.

\begin{figure}[tbp]
\centering
\includegraphics[scale=1]{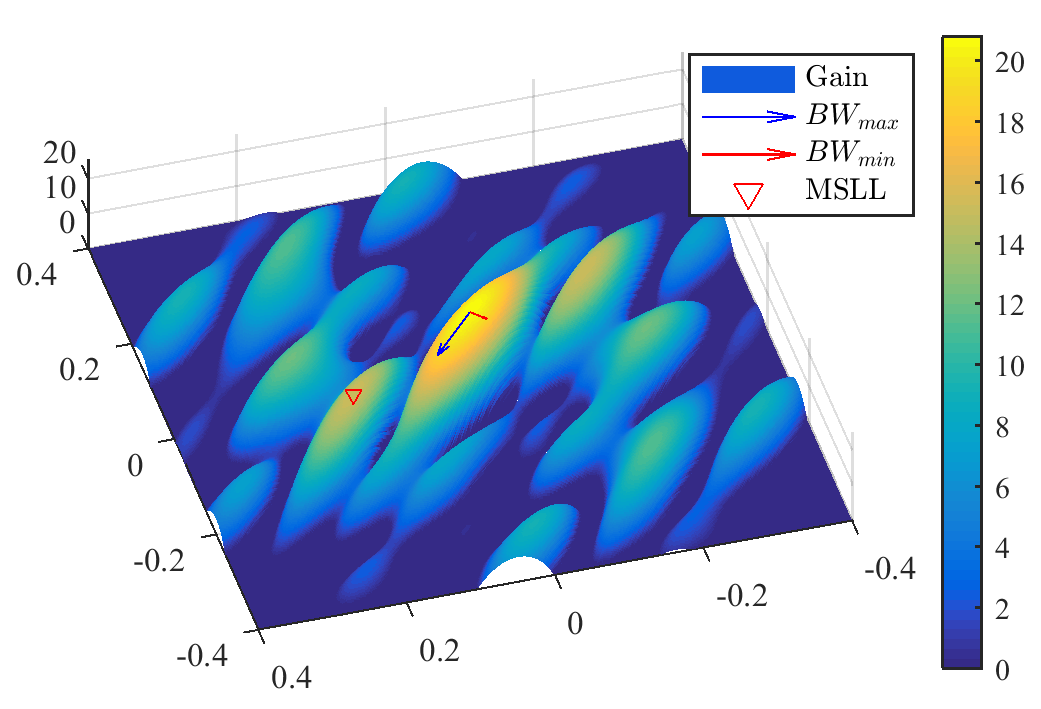}
\caption{Beam Attributes from array Beampattern.}
\label{perf_metrics}
\end{figure}

When beamforming in a general direction $(u_0,v_0)$ (broadside corresponds to $(u_0,v_0) = (0,0)$), the beam pattern is simply given by 
\begin{equation}
\label{beampattern_eq2}
R_{(u_0,v_0)} (u,v) = R(u-u_0,v-v_0)
\end{equation}
For Direction of Arrival (DoA) estimation, the ideal beam should have small beamwidth with minimal sidelobes and high directivity. 
We consider the following beam attributes, some of which depend on the steering direction $(u_0,v_0)$, as key performance metrics to be optimized:
\begin{itemize}
\item  \emph{2D beamwidth} (\texttt{BW}): Although the mainbeam of non-uniform array has non-trivial shape in 2D, we approximate it as an ellipse to define beamwidth. We evaluate the 2D beamwidth in terms of the 3-dB beamwidths along the major and minor axes of this ellipse, denoted by $\mathrm{BW_{max}}$ and $\mathrm{BW_{min}}$, respectively. 
The mean squared error of DoA estimation depends on the sum of these beamwidths (see Appendix \ref{DoA_BW}), hence we define beamwidth as 
$\mathrm{BW^{DoA}=\sqrt{BW_{max}^2+BW_{min}^2}}$.
\item \emph{Maximum sidelobe level} (\texttt{MSLL}):  is the relative level of the strongest sidelobe in the beampattern with respect to mainlobe. $\mathtt{MSLL}=10\log\left(\mathrm{R_{max}/R_{sidelobe}}\right)$.
$R_{\mathrm{max}}, R_{\mathrm{sidelobe}}$ are the largest and second largest magnitude local maxima's of beampattern $R(u,v)$ given by,
\begin{align*}
R_{\mathrm{max}} (u_0,v_0) =& \max_{u,v} R_{u_0,v_0} (u,v) \\
=& R_{u_0,v_0} (u_0,v_0) = R(0,0)\\
R_{\mathrm{sidelobe}}=& \max_{u^*,v^*}R_{u_0,v_0} (u^*,v^*)  \\
& \text{s.t.}\: (u^*,v^*) \neq (u_0,v_0),\\
& \quad \: R_{u_0,v_0} (u^*,v^*)\geq R_{u_0,v_0} (D_{\epsilon}(u^*,v^*)) 
\end{align*}
where $D_{\epsilon}(u^*,v^*)=\left\lbrace (u,v):|u-u^*|<\epsilon, |v-v^*|<\epsilon \right\rbrace$ denotes $\epsilon$-neighborhood. Note that $R_{\mathrm{max}}$ does not depend on steering direction
$(u_0,v_0)$, but $R_{\mathrm{sidelobe}}$ might.
\item \emph{Directivity} ($\mathtt{G_D}$): The directivity is the ratio of mainlobe power to average power, $\mathtt{G_D}=10\log\left( \mathrm{\frac{R_{max}}{R_{avg}}}\right)$
The average power does not have closed form expression for general planar arrays and is evaluated in $(u,v)$ domain by the integral \cite{nuttall2001approximations},
\begin{align*}
R_{avg}=\frac{2}{4\pi}\int_{-1}^1\int_{-\sqrt{1-v^2}}^{\sqrt{1-v^2}}\frac{R_{u_0,v_0}(u,v)}{\sqrt{1-u^2-v^2}}dudv
\end{align*}
\item \emph{Eccentricity} (\texttt{ecc}): is a measure of the asymmetry of the mainbeam. We add this additional parameter to suppress the trivial linear placement solution, $\mathtt{ecc} =\sqrt{1-\left(\mathrm{BW_{min}/BW_{max}}\right)^2}$ 
\end{itemize}
For non-uniform planar arrays, none of these beam parameters have a closed form expression, hence they must be computed numerically. In our simulations, we compute these beam attributes using a beampattern over a $512 \times 512$ grid in UV space as shown in Figure \ref{perf_metrics}. 
\subsection{Problem Formulation}

\begin{figure*}[htbp]
\centering
\includegraphics{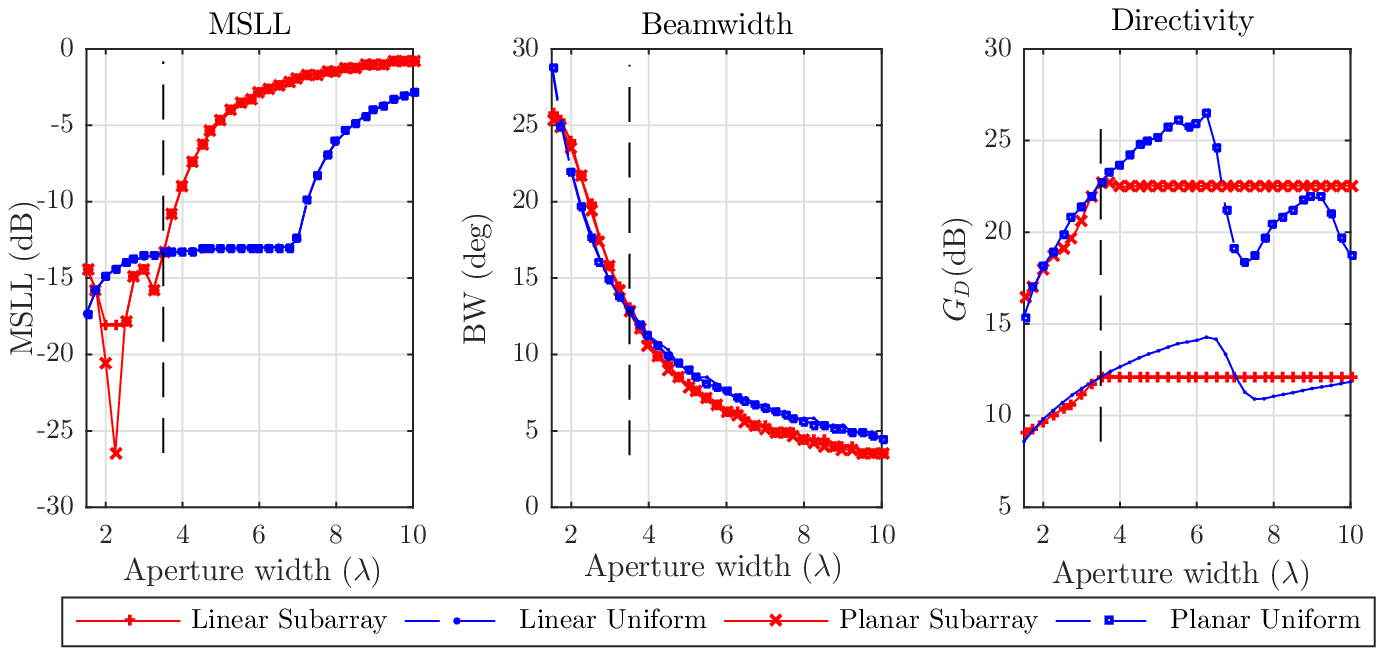}
\hspace{0.5 cm}
\includegraphics{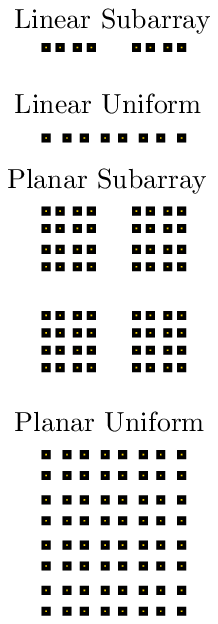}
\caption{Comparison of beam attributes of subarrayed and uniform architecture with increasing aperture width.}
\label{directivity_increase}
\end{figure*}

In order to develop geometric heuristics for optimization, we first analyze the effect of increasing aperture width, (keeping the number of antenna elements fixed, for linear and planar arrays, with uniform and subarray-based architectures as shown in the rightmost section of Figure \ref{directivity_increase}. In the plots of beam attributes in Figure \ref{directivity_increase}, the dashed line represents the aperture width when inter-element spacing equals a half-wavelength, when the uniform and subarray-based configurations match. 

\begin{itemize}
\item The MSLL for subarrayed array increases much faster than uniform configuration due to grating lobe appearing close to mainbeam. This attribute is sensitive to element distribution and behaves unpredictably for non-uniform arrays.
\item The 3dB Beamwidth ({$\mathtt{BW}^{max}$} for planar array) for both array types reduces congruently reaffirming that it is inversely proportional to aperture width independent of the distribution of elements. 
\item Directivity increases as we increase the inter-element spacing, but only upto certain limit and then becomes constant \cite{lee2000evaluation}. This generalizes well to planar arrays as shown in the Figure \ref{directivity_increase}. As one can see the directivity for subarrayed configurations remains approximately constant with increasing aperture width beyond standard spacing. Hence we do not include this metric in our cost function.
\end{itemize}

The objectives that we wish to trade off against each other do not have the same units: for example, \texttt{MSLL} is measured in dB from max sidelobe level, whereas \texttt{BW} is measured in $\mathrm{deg}$.
We therefore normalize each raw objective value, $o^\mathrm{raw}$ by its range as follows:
\begin{align*}
o(\bm{C})=\frac{o^\mathrm{raw}(\bm{C})-\underset{\forall\bm{C}}{\min}\left\{o^\mathrm{raw}(\bm{C})\right\}}{\underset{\forall\bm{C}}{\max}\left\{o^\mathrm{raw}(\bm{C})\right\}-\underset{\forall\bm{C}}{\min}\left\{o^\mathrm{raw}(\bm{C})\right\}}
\end{align*}
The range of each objective is computed numerically while constructing the dictionary of all configurations, $\bm{C}\in \mathscr{C}$, described later in \ref{placement_search}.

The constrained multi-objective optimization can now be formulated as follows:
\begin{equation}
\label{cost_general}
\begin{aligned}
\bm{C}^*,\bm{w}^*=&  \text{arg}\underset{\forall \bm{C,w}}{\min} &&  f(\bm{C,w})\\
& \text{subject to} &&  \textsc{AoS}(\bm{C})
\end{aligned}
\end{equation}
where
\begin{equation}
\label{cost_function}
f(\bm{C,w})=\alpha\mathtt{BW}(\bm{C,w}) + \beta\mathtt{MSLL}(\bm{C,w}) + \gamma \mathtt{ecc}(\bm{C,w})
\end{equation}
is the weighted cost function in terms of the normalized objective functions and [$\alpha,\beta,\gamma$] are weights that can be used to sweep through the optimal surface for this optimization. 
We show some example arrays obtained for different choices of weights in Table \ref{table_example}. 

This is a combinatorial optimization problem which is non-convex, with non-differentiable geometric constraints. The objective function can be evaluated exactly given the subarray centers, but
exploring the entire solution space is computationally infeasible.  Furthermore, beam characteristics in general depend on the steering weights, which in turn depend on the direction $(u_0,v_0)$
in which we are steering.
We therefore employ two key simplifications:
\begin{itemize}
\item We remove the dependence of the cost function on beamforming direction, and hence on steering weights, by computing the objectives based on an expanded
beam pattern, as discussed in Section \ref{sec:invariance}.
\item We employ geometric heuristics to cut down the solution space to a reasonable size, as described in Section \ref{placement_opt}. 
\end{itemize}

\subsection{Invariance to Beamforming Direction} \label{sec:invariance}

The cost function in \eqref{cost_function} is evaluated using the beampattern $R(u-u_0,v-v_0)$, which depends on the beamsteering direction ($u_0,v_0$).
It would be prohibitively expensive to evaluate the beam attributes over all such beampatterns for finding the optimal ($\bm{C,w}$). However, for arrays with isotropic elements, we can
define an Expanded Beam pattern (EBP) which subsumes beampatterns of all steering direction in a Region of Interest (ROI) \cite{lange2011antenna}. 
Suppose that our maximum steering angle in the ROI is $\theta_{max}$. From (\ref{uv_defn}), we see that $(u_0,v_0)$ lies within a circle of radius $\sin \theta_{max}$.
On the other hand, sidelobes can appear at any $(u,v)$ within a circle of radius $1$.  It is easy to see, therefore, that $(u-u_0,v-v_0)$ is guaranteed to lie within a circle
of radius $1 + \sin \theta_{max}$.  We can therefore compute beam attributes using the following EBP:
\begin{equation}
\label{EBP}
\begin{aligned} 
R_\rho(\tilde{u},\tilde{v})&=\frac{1}{N^2}\left |\sum_{i=1}^N {e^{jk\rho\left( \tilde{u}d_i^x+\tilde{v}d_i^y \right)}}\right |^2\\
\rho&=1+\sin(\theta_{max})
\end{aligned}
\end{equation}

\begin{figure}[htbp]
\centering
\includegraphics[scale=1]{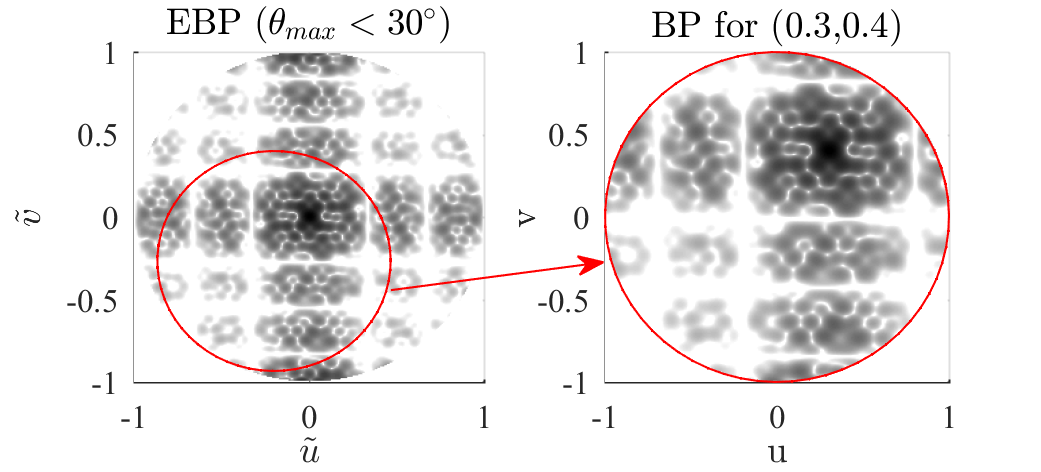}
\caption{Expanded Beampattern for ROI ($\theta_{max}<30^\circ$)}
\label{MBP}
\end{figure}

Figure \ref{MBP} shows the EBP, $R_{1.5}(\tilde{u},\tilde{v})$ for ROI with $\theta_{max} = 30^\circ$, and the beampattern for the steering angle ($(u_0,v_0)=(0.3,0.4)$). The shape of the main beam
is preserved under the transformation \eqref{EBP}, hence beam width and eccentricity can be directly evaluated from \eqref{EBP}. The \texttt{MSLL} evaluated from EBP is a worst-case value, 
corresponding to an argument $\rho (\tilde{u},\tilde{v})$ which, in principle, might not correspond to a feasible value of $(u-u_0,v-v_0)$ in (\ref{beampattern_eq2}). However, the maximum sidelobe always lies within the main lobe of the subarray beam pattern, so that physically implausible values of $\rho (\tilde{u},\tilde{v})$ do not correspond to large local maxima of the EBP.

%

With the introduction of the EBP, we can, without loss of generality, assume that the main beam is being steered towards broadside, setting $\bm{w}=\mathbbm{1}_N$. Our problem now reduces to finding the optimal configuration $C^*$.

%

\section{Placement Optimization}
\label{placement_opt}
In order to optimize the placement, we need to evaluate the cost function over all array configurations. The number of configurations depends on the allowed form factor, and the size of the
subarray module, and an exhaustive search over all configurations is computationally infeasible: for example, the number of configurations for a discrete grid of size $20\times20$ is of the order $10^{20}$.
We therefore propose a two-stage approach, first performing a combinatorial search on a reduced search space, and then obtaining the final solution by searching over perturbations around the
solution from the first stage.

\subsection{Combinatorial search}
\label{placement_search}
\begin{figure}[htbp]
\centering
\includegraphics[scale=1]{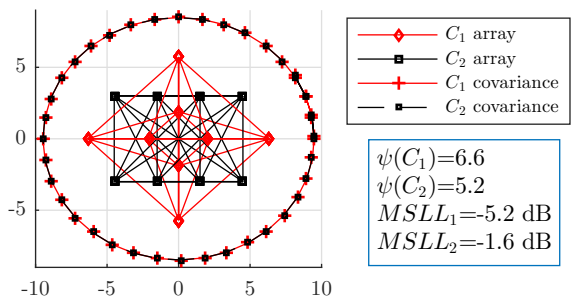}
\caption{Super-arrays with equal covariance but different beam attributes.}
\label{discrimination}
\end{figure}

We reduce the solution space by removing geometrically ``similar'' arrays. We employ the covariance of the element positions, and pairwise element separations, as measures of similarity.
The choice of covariance of element positions $\bm{\Sigma}(\bm{C})$ as similarity metric is motivated by its inverse proportionality to \CRB on accuracy of DoA estimation (see Section \ref{CRB}). However, array configurations with similar array covariance but diverse beam attributes also exist: Figure \ref{discrimination} shows an example of two array configurations with different shapes but the same covariance. The arrays have similar beamwidth but their \texttt{MSLL} levels are different. We observe that the variance of pairwise element distances $\psi(C)$ is different for these arrays, and
use it as an indicator for these large scale deviations.
This allows us to reduce the dimensionality of the solution space from $2\times N_{sub}$ down to $3$ array shape parameters: the eigenvalues ($\lambda_1,\lambda_2$) of $\bm{\Sigma}(\bm{C})$ and $\psi(C)$.


\subsubsection{Subarray Placement Algorithm}
We construct a prefix tree dictionary to find feasible solutions using a breadth first search based enumeration technique.  
The element position covariance for an array of subarray can be uniquely represented by the covariance of its subarray centers, $\bm{\Sigma}_D=\bm{\Sigma}_C+\bm{\Sigma}_{D_e}$. Hence the super-array center covariance can be used instead of that of the full array in the Dictionary search algorithm.
Each node in the tree stores a subarray center position, and the path from root to a node at the $n^{th}$ layer of prefix tree represents a unique configuration of $n$ subarrays. 
The subarray centers are constrained to lie on a fixed set of discrete grid points $G$. 
\begin{algorithm}
\caption{Prefix Tree Dictionary Search}
\label{prefixtreealgorithm}
\begin{algorithmic}[1]
\STATE {\textsc{Initialize}: $\mathscr{C}^1=\left\lbrace C_i^{(n=1)}\right\rbrace; i\in\left[1, N_{init}\right]$}
\WHILE{$n<N_{s}$}
	\STATE \textsc{List} all vacant Gridpoints $V_i=\mathscr{T}_G(C_i^n);i\in\left[1,|\mathscr{C}^n|\right]$
	\STATE \textsc{Append} subarray at vacancies $V_i$, \\$\hat{\mathscr{C}}=\bigcup\limits_{i=1}^{|\mathscr{C}^n|} C_i^n \times V_i$
	\STATE \textsc{Prune}: $\mathscr{C}^{n+1}\leftarrow$ \textcolor{red}{Prune} $\left(\hat{\mathscr{C}}\right)$
	\STATE $n=n+1$
\ENDWHILE
\STATE {Return $\mathscr{C}^{N_{s}}$}
\end{algorithmic}
\end{algorithm}%
The algorithm is described in Algorithm \ref{prefixtreealgorithm}. We briefly discuss the key steps below.
\begin{itemize}
\item \textsc{Initialize}: In order to allow for sufficient exploration, we employ multiple random initializations $C_i^1$  of the root node being placed on $N_\mathrm{init}$ different locations on the grid. (For example, circular configurations cannot be obtained if the root subarray is fixed at the center.)
\item \textsc{List}: Define the operator $\mathscr{T}_G:G^{|C_i^n|}\rightarrow G^{|V_i|}$ which maps the set of subarrays centers $C_i^n \in \mathscr{C}^n$ to the set of $|V_i|$ vacant gridpoints in $G$ available for placement of next subarray which are not blocked by the subarrays already placed at $C_i^n$. This operator also accounts for additional surface area occupied by the subarray module apart from the physical antenna elements (see Appendix \ref{vacancy} for details). 
\item \textsc{Append}: The $(n+1)^{th}$ subarray configuration is constructed from the vacancies, $C_i^n \times V_i$ where $\times$ denotes cartesian product of sets. A temporary dictionary $ \hat{\mathscr{C}}$ is formed by inserting $|V_i|=\kappa$ child nodes for each node in the $n^{th}$ layer.  
\item \textsc{Prune}: Nodes corresponding to ``similar'' configurations are deleted based on the array shape parameters
\begin{enumerate} 
\item Find eigenvalues ($\lambda_1,\lambda_2$) of the subarray center covariance matrix, $\Sigma(C)$ and variance of array separations, $\psi(C)=\Ebb\left[(l_{ij}-\Ebb[l_{ij}])^2 \right]$, where, $l_{ij}$ denotes the distances betweeen $i^{th}$ and $j^{th}$ elements. 
\item Enumerate unique configurations by binning the ($\lambda_1,\lambda_2,\psi$) triplets over a 3-D grid with resolution $\tau$ and randomly picking one configuration from each bin (see Appendix \ref{pertb} for criteria to choose $\tau$).
\end{enumerate}
\end{itemize}
The procedure is repeated until dictionary atoms reach the desired number of subarrays i.e $n=N_S$.
 All arrays in the dictionary $\mathscr{C}$ obtained from this algorithm satisfy the \textsc{AoS}$(\bm{C})$ constraint by construction. This simplifies the constrained Multi-objective optimization in Eq. \eqref{cost_general} to 
\begin{equation}
\label{costA}
\begin{aligned}
\bm{C}^*=\text{arg}\underset{\bm{C}\in\mathscr{C}}{\text{min}} f(\bm{C},\mathbbm{1}_N)
\end{aligned}
\end{equation}

\subsection{Iterative Placement Refinement}

In the second stage, we try to improve the cost function (\ref{costA}) by applying small local perturbations (within a bin of the grid $G$) to the subarray positions obtained from the combinatorial search in the first stage, as described in Algorithm \ref{local_refinements}.

\begin{algorithm}
\caption{Local Refinements}
\label{local_refinements}
\begin{algorithmic}[1]
\STATE {\textsc{Initialize}: $C=C^{init}$; $B$ = oversampled bin.}
\WHILE{$n<N_{ref}$}
	\FOR{$i=1$ \TO $N_s$}
		\STATE \textsc{List} Find positions available for adjustment, $V_i=\mathscr{T}_B(C\setminus C_i)$
		\STATE \textsc{Correct}: Select position with least cost ${C_i}\leftarrow\min_{\bm{b}\in V_i}f\left(\left\{(C\setminus C_i),\bm{b}\right\},\mathbbm{1}\right)$
	\ENDFOR
	\STATE $n=n+1$
\ENDWHILE
\STATE {Return $C$}
\end{algorithmic}
\end{algorithm}%

\begin{figure}[htbp]
\centering
\includegraphics[scale=0.5]{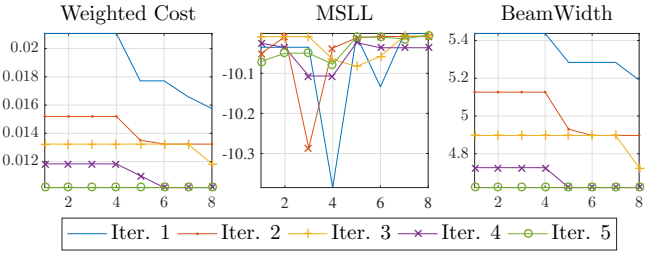}
\caption{Objective costs variation over iterative refinements.}
\label{cost_refinements}
\end{figure}
Figure \ref{cost_refinements} shows a sample of how costs are minimized using sequential refinement. After running few iterations, the final array has $0.8^\circ$ lower beamwidth, while keeping other beam attributes relatively unchanged. 


\section{Estimation-Theoretic Benchmarks}
\label{estimation}

We now seek to evaluate the efficacy of our sparse designs for the canonical application of 2D DoA estimation.  We compare different array designs in terms of estimation-theoretic bounds
as well as simulated performance using a super-resolution algorithm. 
For clarity in exposition, from here onwards we overload $\bm{u} \triangleq [u,v]$ to denote the DoA.

\subsection{Signal Model}

We model the received signal from $K$ sources in the scene with distinct DoAs $\Theta =[\bm{u_1,u_2,\cdots,u_k}]$ as
\begin{align}
\label{rx_signal}
\bm{x}=\sum_{j=1}^K \alpha_j \bm{s}(\bm{u}_j)+\bm{z}
\end{align}
where $\bm{s(u_j)}=\left[e^{jk\bm{u_j^Td_1}}, \hdots, e^{jk\bm{u_j^Td_N}}\right]^T$ is the array response, $\bm{z}=[z_1, \hdots ,z_N]^T$ is complex white noise such that $\mathbb{E}(zz^H)=\sigma^2I_N$, and $\{ \alpha\}_{j=1}^K$ are complex gains which are unknown deterministic constants. The joint probability density of received signal conditioned on $(\Theta, \{ \alpha\}_{j=1}^K)$ is given by,
\begin{align}
p(\bm{x}|\Theta,\bm{\alpha})=\prod_{\bm{u}_j\in\Theta}\frac{1}{\pi^N\sigma^{2}}\exp\left(-\frac{\norm{\bm{x}-\alpha_j\bm{s}(\bm{u}_j)}^2}{\sigma^2} \right) \label{joint_pdf}
\end{align}
For any DoA estimator $\hat{\Theta}$, the covariance of estimation error is defined as,
\begin{align*}
\bm{R}_\epsilon(\hat{\Theta})=\mathbb{E}\left[\sum_{i=1}^K (\bm{u-\hat{u}_{i}})(\bm{u-\hat{u}_{i}})^T\right]
\end{align*}
$\bm{R}_\epsilon$ can be geometrically interpreted by its trace $\sqrt{tr(R_\epsilon)}$ which represents the expected overall Root mean square error (RMSE) in DoA estimation (see Appendix \ref{DoA_BW}). We use this measure to compare the performance of array designs in Section \ref{results_all}.
For single source case $(K=1)$, the joint maximum likelihood estimator of $\bm{u}$ and $\alpha$ yields a noncoherent estimator for $\bm{u}$
as follows:
\begin{align}
\label{MLE}
\bm{\hat{u}}_{ML}=arg\max_{\bm{u}} \left| \bm{s(u)^H x} \right|^2 
\end{align}
For this case, we derive the Cramer Rao (CRB) and Ziv-Zakai (ZZB) bounds on $\bm{R}_\epsilon$ to assess the best possible estimation accuracy of different designs.
Although derived for single source case, we use these bounds for multiple source case as well to compare DoA estimation performance.
 \subsubsection{\CRB}
 \label{CRB}
The Bayesian \CRB for this signal model is given by \cite{lange2011antenna}:
\begin{align}
\label{CRB_eq}
CRB(\bm{R_\epsilon})=(\bm{J_F} + \bm{J_P})^{-1}
\end{align}
where $\bm{J_F, J_P}$ denote the Fisher Information Matrix (FIM) contributions from the observation and the prior distribution of DoA respectively. 
  \begin{align*}
  (\bm{J_F})_{ij}=-\Ebb_{\bm{x,u}}\left[\frac{\partial^2 l(\bm{x|u})}{\partial u_i \partial u_j}\right], 
  (\bm{J_P})_{ij}=-\mathbb{E}_{\bm{u}}\left[\frac{\partial^2 l(\bm{u})}{\partial u_i \partial u_j}\right]
  \end{align*}
where $l(\bm{x|u}), l(\bm{u})$ are the conditional log likelihood and prior log likelihoods respectively. In addition the following regularity condition needs to be satisfied,
\begin{align*}
\Ebb_{\bm{x,u}}\left[\frac{\partial l(\bm{x|u})}{\partial \bm{u}}\right]=\bm{0}\\
\Ebb_{\bm{x,u}}\left[jk\sum_{i=1}^N {\bm{d}_i\left( x_ie^{jk\bm{u}^T\bm{d}_i}-x_i^*e^{-jk\bm{u}^T\bm{d}_i}\right)} \right]=\bm{0}\\
jk\left(\alpha\sum_{i=1}^N {\bm{d}_i} -\alpha^*\sum_{i=1}^N {\bm{d}_i}\right)=jk(\alpha-\alpha^*)\sum_{i=1}^N {\bm{d}_i}&=\bm{0} 
\end{align*}
In order to always satisfy this condition, we enforce the array element positions to be centered i.e. $\sum_{i=1}^N {\bm{d}_i}=\bm{0}$

For single source, the FIM is given by,
  \begin{align}
  J_F&=-\frac{1}{\sigma^2}\Ebb\left[\frac{\partial{\bm{s(u)}}}{\partial \bm{u}}^H\frac{\partial{\bm{s(u)}}}{\partial \bm{u}}\right]\label{FIM}\\
  &=2k^2\gamma\bm{D}^T\bm{D}
  \end{align}
which depends only on the element positions, $D$ and Signal to Noise ratio (SNR) ($\gamma=|\alpha|^2/\sigma^2$).
 Assuming the DoA prior to be uniformly distributed in the ROI ($\theta \leq 30^{\circ} $), the prior FIM simplifies to $\bm{J_P}=1.343\bm{I}_2$.

\subsubsection{Ziv-Zakai Bound}

The CRB is a local bound, which accounts for estimation performance dependent on mainbeam, hence it is only useful at high SNR. In order to better characterize the estimation performance of Sparse arrays at low SNR, we calculate the Ziv-Zakai Bound (ZZB) which incorporates the effect of sidelobes and predicts the threshold behavior. 
For any directional vector $\bm{a}=[\cos\xi,\sin\xi]^T$, the ZZB is given by \cite{bell1996explicit} 
\begin{align*}
\boldsymbol{a^TR_\epsilon a}\geq\int_0^\infty{\mathcal{V}\left\{\max_{\boldsymbol{\delta:a^T\delta}=h}\int A(\bm{u,\delta})P_e(\bm{u,\delta})d\bm{u}\right\}hdh}
\end{align*}
where, $A(\bm{u,\delta})=\min{\left\{p(\bm{u}),p(\bm{u+\delta})\right\}}$, $\mathcal{V}(.)$ is the valley filling function and $P_e(\boldsymbol{u,\delta})$ is error probability of the following vector parameter binary detection problem,
\begin{align*}
H_0&: \boldsymbol{\hat{u}=u};\quad Pr(H_0)=\frac{1}{2},\space \bm{x}\sim p(\bm{x}|\pmb{u})\\
H_1&: \boldsymbol{\hat{u}=u+\delta};\quad Pr(H_1)=\frac{1}{2},\space \bm{x}\sim p(\bm{x}|\pmb{u+\delta})
\end{align*}
This error probability can be lower bounded by the minimum probability of error of the following optimal non-coherent detector:
\begin{align*}
\mathrm{Decide}(\bm{u})=\begin{cases}
      \bm{u} & \text{if } \rho_1 > \rho_2\\
      \bm{u+\delta} & \text{if } \rho_1 < \rho_2
    \end{cases}    
\end{align*}
where $\rho_1=|\boldsymbol{x^Hs(u)}|$, $\rho_2=|\boldsymbol{x^Hs(u+\delta)}|$. Given $\bm{u=u_0}$, $\rho_1,\rho_2$ are rician distributed with scale parameter $s=\sigma^2/M$ and non-centrality parameter $\nu=|\alpha| N, |\alpha R(\bm{\delta})| N$ respectively where $R(\bm{\delta})=R(\delta_x,\delta_y)$ is the beampattern from \eqref{beampattern_eq}. The error probability is given by \cite{salehi2008digital}
\begin{align}
P_{nc}(\boldsymbol{u,\delta})&=\frac{1}{2}\left( Pr\left(\rho_1 < \rho_2 |\bm{u}\right)+Pr\left(\rho_1 > \rho_2 |\bm{u+\delta}\right)\right) \nonumber\\ 
&=Pr\left(\rho_1 < \rho_2 |\bm{u}\right)\nonumber\\ 
&=Q_1(a,b)-\frac{1}{2}e^{-\frac{a^2+b^2}{2}}I_0(ab) \label{zzbPe}\\
\text{where,}\nonumber\\\nonumber
a&=\sqrt{\frac{\gamma N}{2}\left( 1-\sqrt{1-|R(\bm{\delta})|^2} \right)}\\\nonumber
b&=\sqrt{\frac{\gamma N}{2}\left( 1+\sqrt{1-|R(\bm{\delta})|^2} \right)}\nonumber
\end{align}
which is not a function of $\bm{u}$. For ROI in our case, the maximum error $h^{\mathrm{(max)}}=\left(\bm{a}^T\bm{\delta}\right)^{\mathrm{(max)}}=1$. Note that for a uniformly distributed DoA in spherical coordinates ($\theta,\phi$), the distribution of $\bm{u}$ is not uniform. However for simplicity of analysis, we make the assumption that $\bm{u}$ is uniformly distributed on a circular disc. Hence, the ZZB expression simplifies to
\begin{align}
\boldsymbol{a^TR_\epsilon a}&\geq\int_0^1{\mathcal{V}\left\{\max_{\boldsymbol{\delta:a^T\delta}=h}\int A(\bm{u})d\bm{u} P_{nc}(\bm{\delta})\right\}hdh} \nonumber\\
ZZB(\boldsymbol{a^TR_\epsilon a})&=\int_0^1{\mathcal{V}\left\{\max_{\boldsymbol{\delta:a^T\delta}=h} P_{nc}(\bm{\delta})\right\}hdh} \label{zzb_eq}
\end{align} 
The maximum error probability over all directions $\bm{\delta}$ cannot be expressed as closed form expression. However, due to the monotonicity of Marcum's Q function, $Q_1(.)$ and Bessel function of $0^{th}$ order, $I_0(.)$, the error probability in \eqref{zzbPe} is maximized only when $R(\bm{\delta})$ is maximized. Therefore, for each values of $h$ we compute the $\max_{\bm{\delta}:\bm{a^T\delta}=h}|R(\bm{\delta})|$ numerically by search over a discrete set of points on the line segment $\bm{a^T\delta}=h$ and substitue in \eqref{zzb_eq}.

\subsection{DoA estimation algorithm}
\label{algo_description}
%
Grid-based sparse estimation for a set of DoAs models the the received signal \eqref{rx_signal} as follows:
\begin{align}
\label{sparse_approx}
\bm{x} =\bm{S(\Psi)}\bm{b}+\bm{z}
\end{align}
where $\bm{S(\Psi)}=\left[\bm{s(u_1)}\cdots \bm{s(u_{|\Psi|})}\right]$ contains the array response at discretized set of DoAs $u_i\in\bm{\Psi}$ as columns. The nonzero entries in $\bm{b}$ point to presence of target in the corresponding DoA in $\Psi$. The DoA and gain pair $(\bm{\hat{u}}_i,\hat{\alpha}_i)_{i=1}^K $ can be estimated by jointly minimizing the residual power,
\begin{align*}
T(\bm{\hat{u}},\hat{\alpha})=\norm{\bm{x}-\sum_{j=1}^K \hat{\alpha}_j \bm{s}(\hat{\bm{u}}_j)}^2
\end{align*}
The NOMP algorithm summarized below provides a two stage estimator:
\begin{enumerate}
\item \textit{Detection}: Using precomputed $\bm{S(\Psi)}$, coarse estimates of DoA and complex gain are obtained
\begin{align*}
\hat{\bm{u}} &= arg\max_{\bm{u\in\Psi}}| \bm{s}(\bm{u})^H\bm{x} |^2 \\
\hat{\alpha} &= \bm{s}(\bm{u})^H\bm{x}/N
\end{align*}
\item \textit{Refinement}: The estimates are refined using the Newton method:
\begin{align}
\hat{\bm{u}}'&=\hat{\bm{u}}- \left(H_\nabla T(\bm{\hat{u}},\hat{\alpha})\right)^{-1}\nabla T(\bm{\hat{u}},\hat{\alpha})\\
\hat{\alpha}' &= \bm{s}(\hat{\bm{u}}')^H\bm{x}/N
\end{align}
where $H_\nabla T$ and $\nabla T$ denote the Hessian and gradient of $T(\bm{u},\alpha)$ with respect to $\bm{u}$ at current estimate $(\bm{\hat{u}},\hat{\alpha})$ (see \cite{marzi2015compressive} for details).
\end{enumerate} 
The algorithm is repeated with the residual signal, $\bm{r}=\bm{x}-\hat{\alpha}'\bm{s(\hat{\bm{u}}')}$ to estimate other DoAs. The \textit{refinement} steps are repeated after each new detection for all DoAs in a cyclic manner for few rounds to improve accuracy.

The algorithm yields $K_{est}$ DoA estimates, with estimation performance degrading when $K_{est}$ does not match the true number of DoAs, $K_{DoA}$. 
Hence, in order to evaluate the arrays independent of such errors, we implement both algorithm where $K=K_{DoA}$ is known.

We also run extensive simulations with another state of the art algorithm, BPDN \cite{van2011sparse}, with default parameters and 5 refinement stages. The computational complexity of BPDN is significantly higher than that of NOMP. The grid $\bm{\Psi}$ needs to be adapted at each iteration for BPDN, depending on the DoAs, whereas it remains fixed for NOMP $\bm{S(\Psi)}$, and can be precomputed, which makes it suitable for faster implementation in large arrays. Since the NOMP algorithm also yields somewhat better estimation accuracy than BPDN, we only present results obtained with NOMP here.



\section{Numerical Results}
\label{results_all}

\subsection{Design of arrays}
\begin{figure*}[htbp]
\centering
\includegraphics[width=\linewidth]{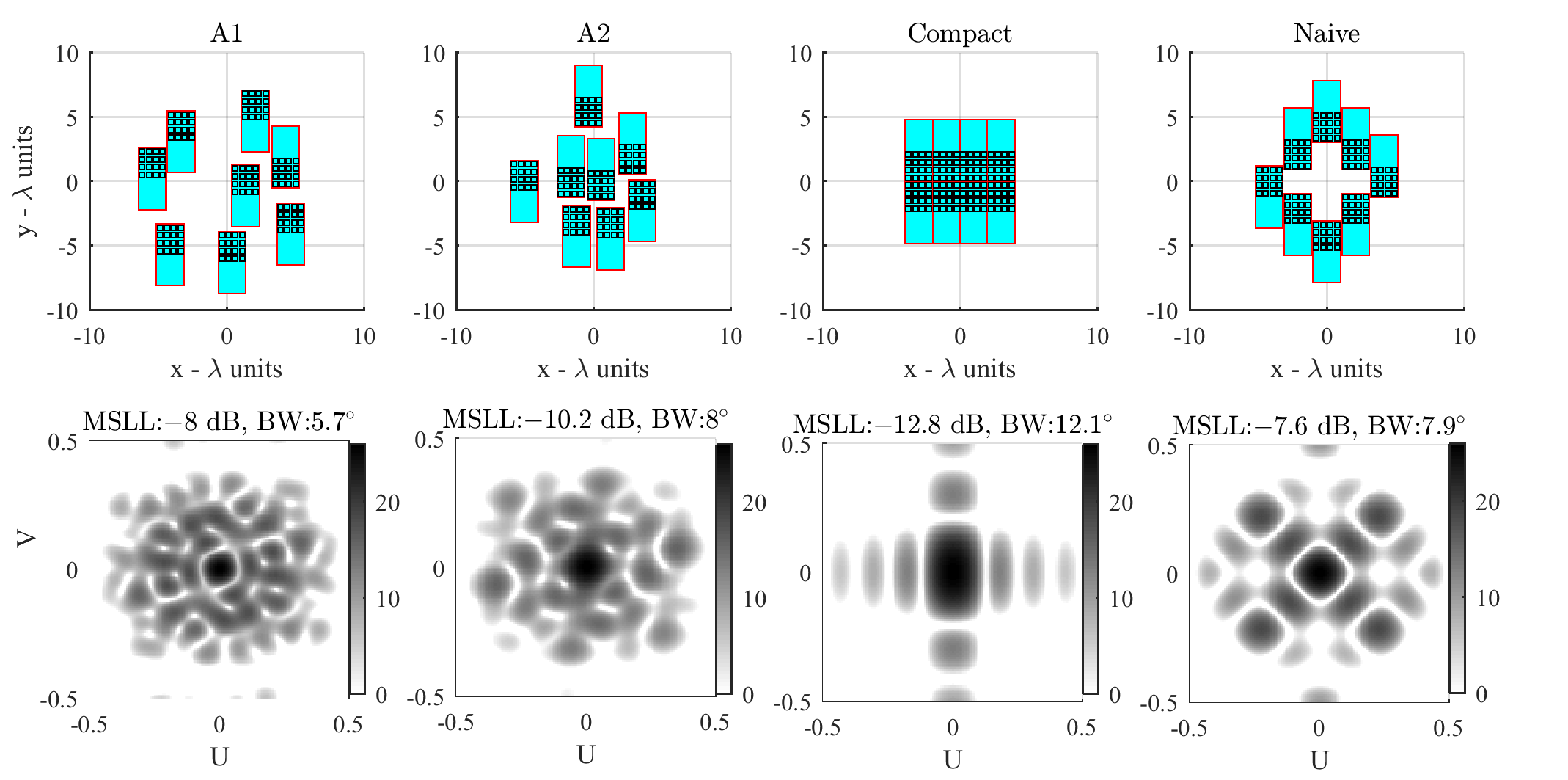}
\caption{Beam patterns (Bottom row) for Designed (Left half) \&  benchmarking (right half) arrays.}
\label{design_vs_naive}
\end{figure*}
Using the combinatorial search algorithm, we create a search space of array configurations, $\bm{C}\in \mathscr{C}$  of size $|\mathscr{C}|= 657,000$ for $N_{sub}=8$ subarrays. We explore this search space to obtain arrays using the placement optimization with various choices of weights given to the cost function in Eq. \eqref{costA}. The following key observations were made about the effect of these weights on subarray placement:  
\begin{itemize}
\item When primary weight is given towards minimizing beamwidth (larger $\alpha$), the subarrays are widely distributed over the available aperture area.
\item When we increase $\beta$ to suppress MSLL, the array becomes restricted to smaller area. 
\item The weight $\gamma$ controls both the shape of main lobe as well as the position of sidelobes. 
\end{itemize}

Based on these observations, we set the weights to obtain following sample array configurations: 
\begin{enumerate}
\item \emph{A1}: Primary emphasis is given towards minimizing beamwidth by setting $\alpha=1$. If multiple arrays are found near the minimum beamwidth, we give secondary priority to other attributes by giving small weight to them ($\beta, \gamma =0.1$).
\item \emph{A2}: In this case, we emphasize all beam attributes by setting all weights equal to 1. 
\end{enumerate}
\begin{table}[htb]
\renewcommand{\arraystretch}{1.3}
\caption{Sample Array Configurations}
\label{table_example}
\centering
\begin{tabular}{|c||c|c|c||c|c|c|}
\hline
Shape & $\alpha$ & $\beta$ & $\gamma$ & \texttt{MSLL} & \texttt{BW} & \texttt{ecc}\\
\hline
\emph{A1} & 1 & 0.1 & 0.1 & -8 dB & $5.7^\circ$ & 0\\
\emph{A2} & 1 & 1& 1 & -10.2 dB & $8^\circ$ & 0\\
\emph{Compact} & - & - & - & -12.8 dB & $12.1^\circ$ & 0.7\\
\emph{Naive} & - & - & - & -7.6 dB & $7.9^\circ$ & 0.0\\
\hline
\end{tabular}
\end{table}
We compare these array designs against two simple array configurations, 1)``compact'' array where subarrays are placed together such that overall element pattern becomes a uniform rectangular array, 2) ``naive'' array where subarrays are spread along a diamond shape such that its resultant beamwidth is equal to that of \emph{A2}. 
Table \ref{table_example} lists the weights and resulting beam attributes of these arrays. 

Figure \ref{design_vs_naive} shows the array designs obtained using our optimization approach and their beam patterns. 
The \emph{A2} array has a sharp beamwidth and only $2.6$ dB worse \texttt{MSLL}  compared to the compact array. On the other hand, a naive sparse array with circular arrangement of subarrays yields $2.6$ dB higher \texttt{MSLL} compared to \emph{A2} for similar beamwidth. Our designs \emph{A1, A2} exhibit several small sidelobes (the highest sidelobe for \emph{A2} is -10.2 dB), 
whereas the naive array exhibits fewer but more pronounced sidelobes. 
Since large sidelobes and grating lobes can cause large errors in DoA estimation, we expect our designs to yield better estimation performance, which is borne out by the results presented in the next section.

\subsection{Comparison of Estimation Performance}
\label{estimation_Results}
We evaluate the arrays based on their DoA estimation accuracy at different SNRs for both single and multiple source cases. We use the RMSE in estimating DoA for comparison which is given by $\bar{\epsilon}=\sqrt{\Ebb[|\bm{\hat{u}-u_0)|^2}]}=\sqrt{tr(\bm{R_\epsilon})/2}$. 
\begin{figure}[htbp]
\centering
\includegraphics[width=0.95\linewidth]{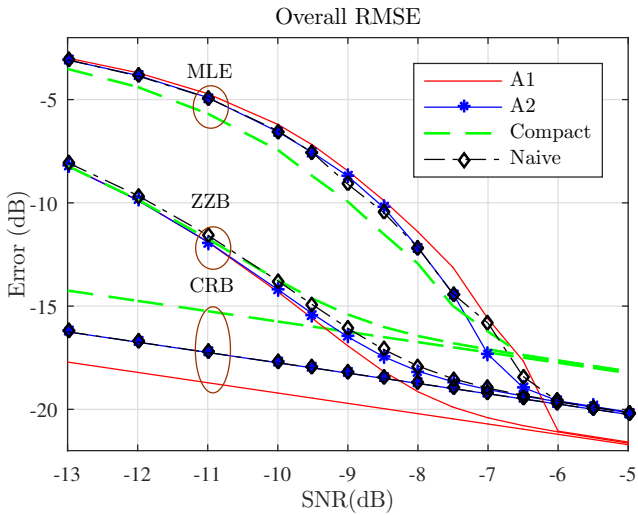}
\caption{Comparison of estimation theoretic bounds for arrays.}
\label{design_vs_naive_RMSE} 
\end{figure}
\subsubsection{Estimation bounds}
The \CRB is evaluated using \eqref{CRB_eq}, $CRB(\bar{\epsilon})=\sqrt{tr(CRB(\bm{R_\epsilon}))/2}$ . The \ZZB is evaluated using \eqref{zzb_eq},
\begin{align*}
ZZB(\bar{\epsilon})=\sqrt{ZZB(\bm{a_1^TR_\epsilon a_1})+ZZB(\bm{a_2^TR_\epsilon a_2})}
\end{align*}
where $\bm{a_1, a_2}$ denote the directions of maximum and minimum beamwidths of the array. We also computed the ML estimation (MLE) error by Monte Carlo simulation using \eqref{MLE} with an overcomplete dictionary of array responses. Figure \ref{design_vs_naive_RMSE} shows the CRB, ZZB and MLE curves for all the arrays. CRB is proportional to beamwidth (see Appendix \ref{BWeval} for details). The ZZB bound converges to CRB at the so-called ``ZZB threshold'' SNR: when the SNR is below this threshold,
far-ambiguities in DoA estimation caused by large sidelobes dominate the MSE. The tradeoff between beamwidth and MSLL is thus expected to translate to one between CRB (better with smaller beamwidth) and ZZB threshold (worse with larger MSLL).    Thus, as expected, ``\emph{A1}'' array achieves the lowest CRB, followed by ``\emph{naive}'' and ``\emph{A2}'' with equal CRB, while ``\emph{Compact}'' array has the largest beamwidth and hence highest CRB. The trend in MSLL is weakly reflected in the ZZB thresholds (for a single target, sidelobes do limited damage):
the degradation in ZZB threshold, relative to that of the compact array for
the optimized arrays (\emph{A1, A2}) is less than $1$ dB, while the gain in CRB due to smaller beamwidth is $4$ dB and $2$ dB, respectively.
The MLE error curve also agrees with the threshold behavior predicted by ZZB.  We see in the next set of results, however, that the size of the sidelobes becomes much more important when we consider multiple targets. 

\subsubsection{Estimation algorithm performance}
We obtain DoA estimates using the \emph{NOMP} algorithm \cite{mamandipoor2016newtonized, marzi2015compressive} with a known number of sources to compare the best case performance of these arrays. 
(The NOMP algorithm also performs as well as the brute force MLE for a single target discussed earlier--results omitted here.)
The RMSE is evaluated across $N=1024$ DoAs uniformly sampled over the ROI (spherical cap of half angle $30^\circ$). 
For evaluating the estimation performance in presence of multiple targets, ($K=5$) we compute the RMS error in the DoA estimate for a primary target fixed at broadside, while interfering targets are distributed uniformly in ROI at separation of $\Delta\bm{u}\geq 0.16$ or $\Delta\theta\geq 9.2^\circ$ away from primary target. 
This separation is imposed because the estimation problem is ill-posed for DoAs in close proximity. 
For uniform arrays, the minimum separation is typically defined with respect to the DFT bin size (e.g. $\mathrm{\Delta_{DFT}}=2\pi/L$ for an $L$-element linear array). 
Since this quantity cannot be defined for non-uniform planar arrays, we choose a minimum separation halfway between the RMS beamwidths of the ``compact'' and sparse arrays,
to capture the effect of both local errors and far ambiguity errors due to sidelobes.  

In addition to RMSE vs SNR curves, we also analyze the distribution of error magnitudes. The complementary cumulative distribution (CCDF) of the estimation errors is used to compare the ``outage probability'' corresponding to too large an error, which captures the impact of large sidelobes. 

\begin{itemize}
\item With multiple sources, the estimation accuracy is degraded by interference from other sources, and RMSE does not converge to the single-target CRB. Fig. \ref{Array_accuracy_multiple} shows the estimation performance for strong and weak interference.
\begin{figure}[htb]
\centering
\includegraphics[scale=1]{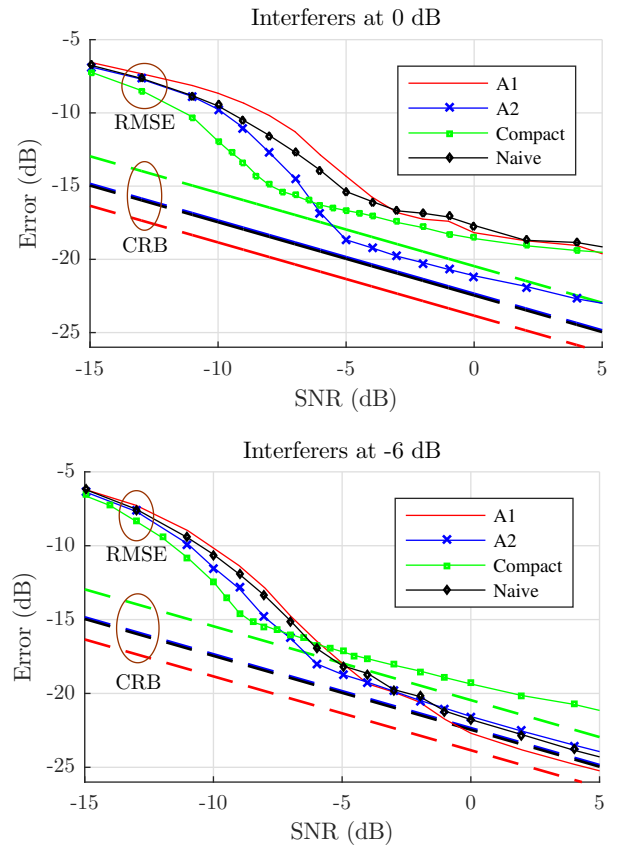}
\caption{Estimation accuracy with multiple targets.}
\label{Array_accuracy_multiple}
\end{figure}
\begin{enumerate}
\item Weak interference: When the interfering sources are 6 dB weaker than primary target, sparse arrays offer more than $5$ dB SNR gain compared to compact arrays for $SNR > -5$ dB. Also, the difference between \emph{A2, naive} array widens to about $1$ dB in the threshold region indicating the benefit of suppressing sidelobes.
\item Strong interference: When interfering sources have same magnitude, RMSE severely degrades for both arrays with high sidelobes (\emph{A1, Naive}) as well as arrays with high beamwidth (\emph{Compact}). On the other hand, \emph{A2} has lowest RMSE at $SNR>-5$ dB because of the dual benefit of small beamwidth and lower sidelobes.
\end{enumerate}

\begin{figure}[htbp]
\centering
\includegraphics[scale=1]{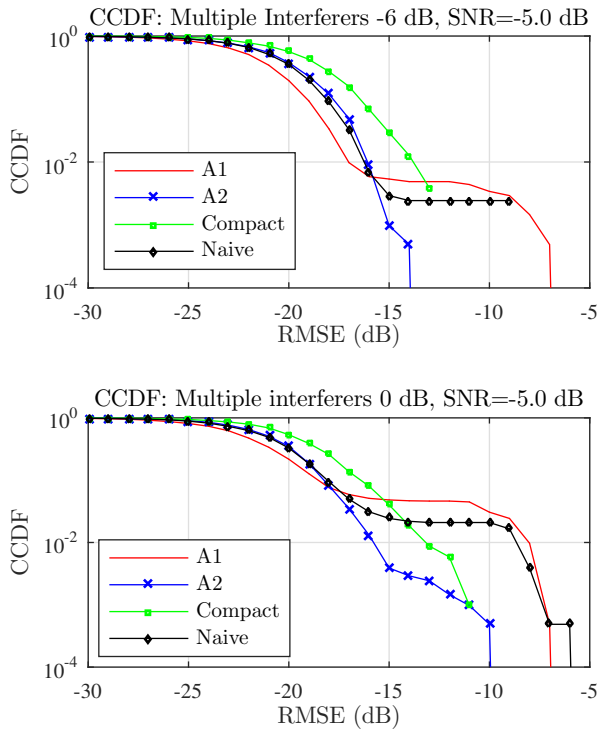}
\caption{CCDF of estimation errors in multiple targets.}
\label{Array_accuracy_CCDF}
\end{figure}
\item The increase in estimation errors at high SNR is attributed to ambiguity errors from sidelobes, hence the overall sidelobe suppression for the arrays can be compared using the distribution of these error magnitudes. Fig \ref{Array_accuracy_CCDF} shows the CCDF curves of all arrays at SNR=$-5$ dB. The initial curvature of these curves (RMSE upto -22 dB) is expected to depend on local errors, hence the rate of change follows same order as CRB which is $\emph{A1} > \emph{A2 = naive} > \emph{compact}$. But the curvature reverses order at higher RMSE indicating the tradeoff with far-errors. We can see that \emph{A2} achieves lower outage probability in both scenarios (e.g. for RMSE threshold set to $-15$ dB) as it strikes a balance between near and far errors. In contrast, both \emph{A1} and \emph{naive} exhibit high outage probability due to frequent far ambiguity errors caused by higher sidelobes.
\end{itemize}
Therefore, depending on the expected magnitude of interferers either one of the designed arrays with suitable sidelobe suppression can be selected. For a desired beamwidth reduction our design algorithm yields an array superior to a naively designed sparse array.

\section{Compressive Estimation}
\label{compressive}
We now evaluate the arrays for sparse estimation using compressive measurements at each subarray given by:
\begin{align*}
\bm{y}=\bm{\Phi x} 
\end{align*}
where $\bm{x}$ is the full measurement from \eqref{sparse_approx} and $\bm{\Phi}=\text{diag}(\bm{\Phi}_1,\cdots,\bm{\Phi}_{N_s})$ is the $M\times N$ measurement matrix consisting of the subarrays measurement matrices as its block diagonals. Each subarray takes $M_i$ compressive measurements  with an independent $\bm{\Phi}_i \in\mathbb{C}^{M_i\times N_e}$ whoose elements are chosen uniformly and independently from QPSK samples $\frac{1}{\sqrt{M_i}}\left\{\pm 1,\pm j\right\}$. In addition, columns of $\bm{\Phi}_i$ have unit norm  to preserve signal norm on average ($\Ebb\left[||\Phi \bm{S(u)}||^2\right] = ||\bm{S(u)}||^2$) while scaling noise variance by $N/M$. The underlying DoA, $\bm{u}$ can be extracted by minimizing the ML cost function:
\begin{align}
&\left\|\bm{y}-\Phi\sum_{j=1}^K\hat{\alpha_j}\bm{s}(\bm{\hat{u}}_j)\right\|^2 \nonumber
\end{align}
The efficacy of compressive parameter estimation in AWGN depends on preserving the geometric structure of the parametrized signals \cite{ramasamy2014compressive}.
Specifically, if $\Phi$ satisfies the $2K$ isometry property for discretized basis $\bm{S(\Psi)}$ \cite{ramasamy2014compressive},
\begin{align}
\label{isometry_eq}
C(1-\epsilon) \leq \frac{|\Phi \bm{S(\Psi)b}|^2}{|\bm{S(\Psi)b}|^2}\leq C(1+\epsilon)
\end{align} 
where $C$ is a constant for any arbitrarily chosen $2K$ sparse vector $\bm{b}$, the performance of the compressive system follows that for the original system, except for an SNR penalty of
$M/N$.
\begin{figure}[htbp]
\centering
\includegraphics[scale=1]{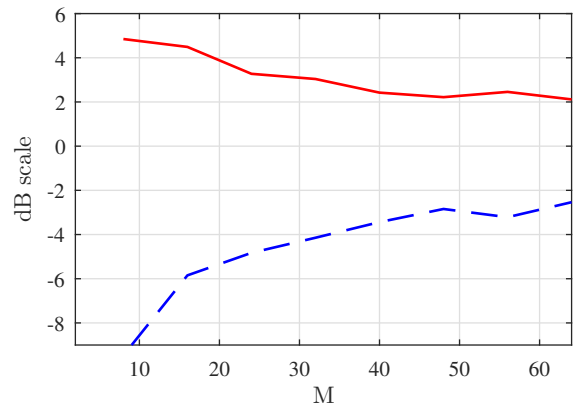}
\caption{Maximum and minimum values of ratio in \eqref{isometry_eq} for sparse array.}
\label{coherence}
\end{figure}
 Figure \ref{coherence} shows the minimum and maximum values of this ratio over $10^6$ random realization of $8$ sparse $\bm{b}$ for sparse array. The ratio is within $[-5,3]$ dB for $M>32$ signifying that $32$ compressive measurements are sufficient to estimate $K=4$ DoAs.

%
\begin{figure}[htbp]
\centering
\includegraphics[scale=1]{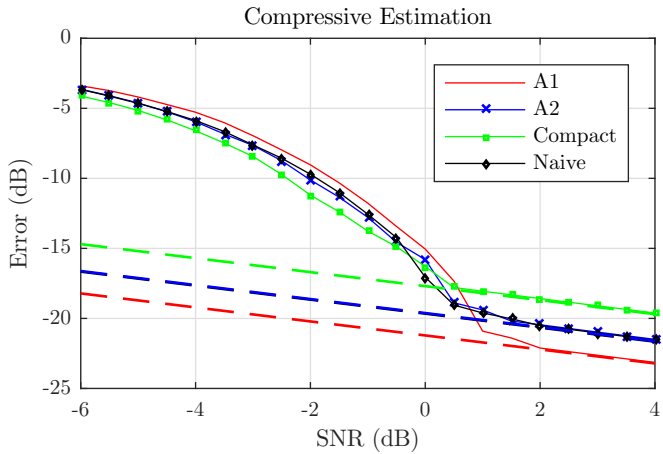}
\caption{Estimation performance with Compressive measurements.}
\label{CS_RMSE}
\end{figure}
Figure \ref{CS_RMSE} shows estimation performance with $M=32$ compressive measurements collected across eight subarrays $(M_i=4, i\in\{1..8\})$.
Comparing with Fig \ref{design_vs_naive_RMSE}, we observe that the estimation algorithms preserve the same characteristics as with full measurements with approximately $6$ dB SNR penalty as expected ($N/M=4$).



\section{Conclusions}

Our results demonstrate that trading off beam width versus side lobes when synthesizing a large effective aperture does indeed produce performance gains in bearing estimation.
Compared to a compact placement of subarrays, an optimized sparse placement produces smaller mean squared error because of its smaller beam width. 
Compared to a naive sparse placement, the control of side lobes via our optimized placement produces the most significant gains when estimating the bearing for multiple sources.
There are a number of interesting directions for further exploration.  It may be possible to improve our optimization framework by alternative approaches for pruning solutions, as well as applying
generic optimization techniques (e.g., genetic optimization) initialized by our solutions.  Exploring the application of our framework for communications, where transmit and receive beamforming gains
are fixed by the number of elements, but control of beam width and sidelobes affects interference, is an interesting direction.  
In the context of sensing, our work may be viewed as design of an individual sensor which can be placed within a more comprehensive architecture, such as a network of sensors for localization
and tracking.

\appendices

\section{Mean square error in 2D DoA estimation}
\label{DoA_BW}
For 2D DoA estimation, the error along any given angle $\xi$ is given by $\bm{a^TR_{\epsilon}a}$, where $\bm{a}=[\cos\xi , \sin \xi]^T$ is the directional cosine and $\bm{R_\epsilon}$ is the error covariance matrix. Assuming that $\{ \nu_i , \bm{q_i} , i=1,2 \}$, denote the eigenvalues and eigenvectors of $\bm{R_{\epsilon}}$, the MSE averaged over $\bm{a}$ (assume $\xi$ uniform over
$[0,2\pi]$) is given by
\begin{align*}
{\rm MSE} &=\Ebb_{\bm{a}}\left[ \bm{a^TR_\epsilon a}\right]  = \nu_1\Ebb_{\bm{a}}\left[\left| \bm{a^T q_1}\right|^2\right]+ \nu_2\Ebb_{\bm{a}}\left[\left| \bm{a^T q_2}\right|^2\right]\\
&=\nu_1 \frac{||\bm{q_1}||^2}{2}+\nu_2 \frac{||\bm{q_2}||^2}{2} = (\nu_1+\nu_2)/2 = \frac{1}{2} tr(\bm{R_\epsilon})
\end{align*}
where we have used $\Ebb [ \cos^2 \xi ] = \Ebb [ \sin^2 \xi ] = \frac{1}{2}$. 



\section{2D Beamwidth \& CRB}
\label{BWeval}
We define 2D beamwidth using the Taylor series expansion of beampattern $R_{\bm{u_o}}(\bm{u})$ around mainlobe $R_{\bm{u_o}}(\bm{0})$. Since the beampattern around the main lobe. and hence
the beamwidth, is invariant to beamforming direction (see \ref{sec:invariance}), we assume $\bm{u_o}=\bm{0}$ without loss of generality, and drop the subscript: $R_{\bm{0}}(\bm{u}) \triangleq R(\bm{u})$. By taking the derivatives of \eqref{beampattern_eq}, theTaylor series expanision up to second order is obtained as
\begin{align}
\label{taylor_exp_simple}
R(\bm{u}) \approx R(\bm{0})-\frac{k^2}{N}\bm{u}^T\bm{D}^T\bm{D}\bm{u}
\end{align}
We define Half Power Beam Contour (\emph{HPBC}) as the closed contour around mainbeam with $\{\bm{u} : R(\bm{u})=0.5 R(\bm{0})\}$, which is approximated as an ellipse using
 \eqref{taylor_exp_simple} as follows:
\begin{align}
\label{hpbc_eq}
\bm{u}^T\bm{D}^T\bm{D}\bm{u}= \frac{N}{2k^2} R(\bm{0})
\end{align}
Consider the eigendecomposition of $\bm{D}^T\bm{D}$ given by,
\begin{align}
\label{eig_dtd}
\bm{D}^T\bm{D}=\lambda_1\bm{p}_1\bm{p}_1^T+\lambda_2\bm{p}_2\bm{p}_2^T \tag{$\lambda_2\geq \lambda_1$}
\end{align}
The eigenvectors $\bm{p}_2,\bm{p}_1$ correspond to major and minor axis of \emph{HPBC} ellipse respectively, and depend only on the element positions. 

\begin{figure}[htbp]
\centering
\includegraphics[width=0.8\linewidth]{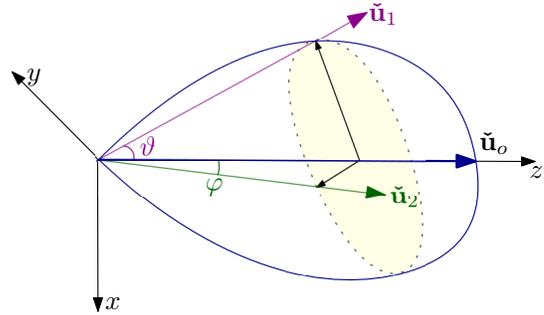}
\caption{2D Beamwidth}
\label{2DBeam}
\end{figure}
Figure \ref{2DBeam} shows the mainlobe of a beam and the dotted shaded region represents its \emph{HPBC} ellipse. 
$ \check{\bm u}_o, \check{\bm u}^1, \check{\bm u}^2$ correspond to unit vectors in the direction of main beam, vertex and co-vertex of the \emph{HPBC} where, $\check{\bm u}=[u,v,\sqrt{1-u^2-v^2}]$ is the unit vector towards directional cosine $\bm{u} = [u,v]$. These can be expressed as,
\begin{align}
\label{ellipse_vertex}
\check{\bm u}_o=\begin{bmatrix}
0\\0\\1
\end{bmatrix}, 
\check{\bm u}_1=\begin{bmatrix}
\sin \vartheta \cos \phi_\mathrm{max}\\ \sin \vartheta \sin \phi_\mathrm{max}\\ \cos\vartheta
\end{bmatrix}, 
\check{\bm u}_2=\begin{bmatrix}
\sin \varphi \cos \phi_\mathrm{min}\\ \sin \varphi \sin \phi_\mathrm{min}\\ \cos \varphi
\end{bmatrix} 
\end{align}
where $\phi_\mathrm{max}, \phi_\mathrm{min}$ are perpendicular azimuthal angles and $\vartheta, \varphi$ are the maximum and minimum beamwidth angles subtended from the major and minor axis of this ellipse to the mainbeam. 
\begin{align}
\label{BW_maxmin}
\mathrm{BW_{max}}=\vartheta=\left(\frac{360}{\pi}\right)\cos^{-1}\left(\bm{\check{u}_o\check{u}_1}\right)\\
\mathrm{BW_{min}}=\varphi=\left(\frac{360}{\pi}\right)\cos^{-1}\left(\bm{\check{u}_o.\check{u}_2}\right)
\end{align}
Substituting the major and minor axis from \eqref{ellipse_vertex} in \eqref{hpbc_eq}, we obtain
 \begin{align*}
\lambda_1\sin^2 \vartheta = \frac{N}{2k^2} R(\bm{0}) \implies \sin \vartheta \approx \mathrm{BW_{max}} \propto 1/\sqrt{\lambda_1}\\
\lambda_2\sin^2 \varphi = \frac{N}{2k^2} R(\bm{0}) \implies \sin \varphi \approx \mathrm{BW_{min}} \propto 1/\sqrt{\lambda_2}
\end{align*}
That is, the beamwidths along extremal directions are inversely proportional to the square roots of the eigenvalues of $\bm{D}^T\bm{D}$.

\subsubsection*{Relation to CRB}
Using \eqref{FIM}, the error covariance matrix is lower bounded by
\begin{align*}
\bm{R_{\epsilon}} \geq \text{CRB} &= J_F^{-1} = \frac{N}{2k^2\gamma}\left(\bm{D}^T\bm{D}\right)^{-1} \\
&=\frac{N}{2k^2\gamma}\left( \frac{1}{\lambda_1}\bm{p}_1\bm{p}_1^T+\frac{1}{\lambda_2}\bm{p}_2\bm{p}_2^T \right) \tag{using \eqref{eig_dtd}}
\end{align*}
Using Appendix \ref{DoA_BW}, the MSE can be lowerbounded by
\begin{align*}
{\rm MSE} \geq \overline{\text{CRB}} = \frac{1}{2} tr(J_F^{-1}) &= \frac{N}{4k^2\gamma}\left( \frac{1}{\lambda_1}+\frac{1}{\lambda_2} \right)\\
&= \frac{N}{4k^2\gamma}\left( \sin^2(\vartheta) + \sin^2(\varphi)  \right)\\
&\propto \frac{\mathrm{(BW^{DoA})^2}}{\mathrm{SNR}} 
\end{align*}
where $\mathrm{BW^{DoA}=\sqrt{BW_{max}^2+BW_{min}^2}=\sqrt{\vartheta^2+\varphi^2}}$ is defined as MSE beamwidth ($\sin \theta \approx \theta$ for small angles $\theta$).

\section{Vacancy search operator $\mathscr{T}$}
\begin{figure}[htbp]
\begin{minipage}{.5\linewidth}
  \centering
\includegraphics[width=0.35\linewidth,angle=270]{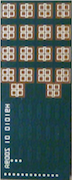}
\end{minipage}%
\begin{minipage}{.5\linewidth}
  \centering
\includegraphics[width=0.35\linewidth,angle=90]{figs/rfem.png}
\end{minipage}%
\caption{The Subarray module and its two possible poses. Golden section are copper patch antennas on the green colored chip.}
\label{rfem}
\end{figure}

Our reference subarray module shown in Fig. \ref{rfem} occupies space in addition to antenna elements. In order to keep element polarizations aligned, these modules can be placed in either \emph{up} ($0^\circ$) or \emph{down} ($180^\circ$) pose. We outline a procedure to list the vacant gridpoints $V_i = \mathscr{T}(C_i^n)$ where the \emph{new} subarray can be placed without overlapping with already placed \emph{dormant} subarrays at $C_i^n$.  We define the subarray state as the center $c$ of the element pattern and its  \emph{pose} $\nu$, since vacancies depend on both parameters.
\begin{align*}
\tilde{c} = \{c,\nu\} \forall c\in C_i^n, V_i
\end{align*}
The pose variable $\nu\in\{\nu^u,\nu^d,\nu^{f}\}$ denotes whether subarray can be placed in up only($\nu^u$), down only($\nu^d$) or free pose ($\nu^{f}$, either up or down) at the location $c$.  
For a given set of \emph{dormant} subarray states, $\tilde{C}_i^n$ we identify all vacant states $\tilde{V}_i$ for placing the \emph{new} subarray. Once a \emph{new} subarray is placed, the 
states of all dormant subarrays are updated (e.g., a free pose may switch to an up pose if the down pose becomes infeasible).

\label{vacancy}

\section{Perturbation of array}
\label{pertb}
In order to design a bin size for pruning array configurations, we analyze the effect of perturbing the location of a single array element on the eigenvalues of the array covariance matrix.
Consider a small perturbation $\bm{\upsilon}=[\upsilon_x,\upsilon_y]$ added to $i^{th}$ array element position: $\bar{\bm{d}}_i=\bm{d}_i+ \bm{\upsilon}$. The covariance for the perturbed array is
\begin{align*}
\bm{\Sigma_{\bar{D}}}&=\left( \bm{D}^T\bm{D}+\bm{\upsilon}^T\bm{\upsilon}+2\bm{d}_i^T\bm{\upsilon} \right)/N = \bm{\Sigma_D}+\bm{\mathcal{G}}+2\bm{\mathcal{H}}
\end{align*}
where
\begin{align*}
\bm{\mathcal{G}}=\frac{1}{N}\begin{bmatrix}
\upsilon_{x}^2 & \upsilon_{x}\upsilon_{y} \\  \upsilon_{x}\upsilon_{y}  & \upsilon_{y}^2\end{bmatrix},\quad
\bm{\mathcal{H}}=
\frac{1}{N}\begin{bmatrix}
\upsilon_{x}d_{xi} & \upsilon_{x}d_{yi} \\  \upsilon_{y}d_{xi}  & \upsilon_{y}d_{yi}\end{bmatrix}
\end{align*}

Using Weyl's inequality \cite{horn1990matrix} for real symmetric matrices, the eigenvalue perturbation is bounded as
\begin{align*}
|\bar{\lambda}_i-\lambda_i|&\leq \norm{\bm{\mathcal{G}}+2\bm{\mathcal{H}}}_2\leq \norm{\bm{\mathcal{G}}}_2+2\norm{\bm{\mathcal{H}}}_2\\
&\leq \norm{\bm{\mathcal{G}}}_F+2\norm{\bm{\mathcal{H}}}_F
\end{align*}
The frobenius norms of $\bm{\mathcal{G, H}}$ are
\begin{align*}
\norm{\bm{\mathcal{G}}}_F=\sqrt{\left(\upsilon_{x}^2+\upsilon_{y}^2\right) } /N\\
\norm{\bm{\mathcal{H}}}_F=2  R_i \sqrt{\left(\upsilon_{x}^2+\upsilon_{y}^2\right) } N
\end{align*}
where $R_i=\sqrt{d_{xi}^2+d_{yi}^2}$ is the distance of the $i^{th}$  element from the array center. 
Hence, the overall variation of eigenvalues with variation $\Delta_e =\sqrt{\left(\upsilon_{xi}^2+\upsilon_{yi}^2\right) } $ of the $i^{th}$ element is 
 \begin{align*}
 |\bar{\lambda}_i-\lambda_i|&\leq \frac{(2R_i+1)}{N}\Delta_e
 \end{align*}
 Thus, the eigenvalues are more sensitive to perturbations in the locations of elements further from the center.
For perturbations within one grid size used in our placement search algorithm, the eigenvalue of the subarray center covariance can vary at most by $\frac{(2R_i+1)}{N_{sub}}\sqrt{2}\Delta_g$, 
and we use this as a guideline for discretizing the eigenvalues for.removing geometrically similar configurations.

\section*{Acknowledgment}
This work was supported in part by Semiconductor Research Corporation (SRC), DARPA, Center for Scientific Computing at UCSB and National Science Foundation under grants CNS-1518812, CNS-1518632 and CNS-0960316.
\ifCLASSOPTIONcaptionsoff
  \newpage
\fi



\bibliographystyle{IEEEtran}
\bibliography{articles}

%

\begin{IEEEbiography}[{\includegraphics[width=1in,height=1.25in,clip,keepaspectratio]{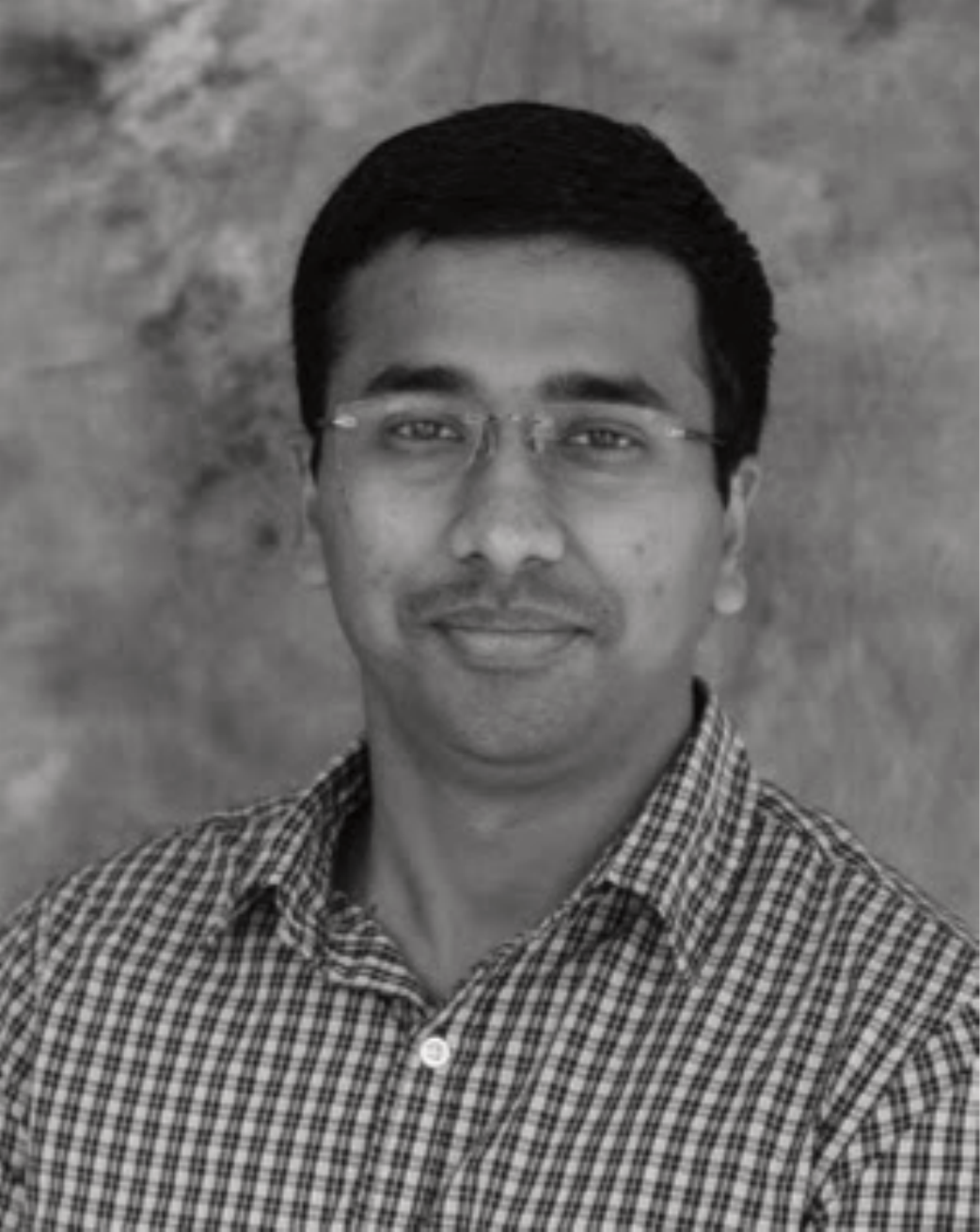}}]{Anant Gupta} received the B.Tech. degree in electronics and electrical communication engineering and the M.Tech. degree in telecommunication systems engineering from IIT Kharagpur in 2013. He received the M.S. degree in electrical and computer engineering from the University of California at Santa Barbara (UCSB) in 2016. He is currently working towards the Ph.D. degree in the Department of Electrical and Computer Engineering, UCSB. His research interests include millimeter wave sensing, statistical signal processing and machine learning.
\end{IEEEbiography}

\begin{IEEEbiography}[{\includegraphics[width=1in,height=1.25in,clip,keepaspectratio]{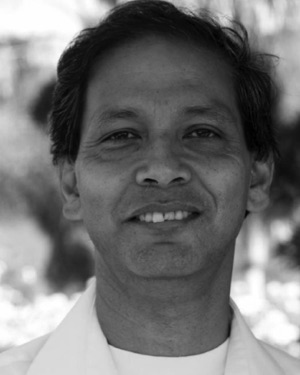}}]{Upamanyu Madhow} received the bachelor’s degree in electrical engineering from IIT Kanpur in 1985 and the Ph.D. degree in electrical engineering from the University of Illinois at Urbana–Champaign, Champaign, IL, USA, in 1990. He was a Research Scientist with Bell Communications Research, Morristown, NJ, USA. He was a Faculty Member with the University of Illinois at Urbana–Champaign. He is currently a Professor of electrical and computer engineering with the University of California at Santa Barbara, Santa Barbara, CA, USA. He has authored two textbooks, Fundamentals of Digital Communication (Cambridge University Press, 2008) and Introduction to Communication Systems (Cambridge University Press, 2014). His current research interests focus on next-generation communication, sensing and inference infrastructures centered around millimeter-wave systems, and on robust machine learning. He was a recipient of the 1996 NSF CAREER Award and a co-recipient of the 2012 IEEE Marconi Prize Paper Award in Wireless Communications. He served as an Associate Editor for the IEEE Transactions on Communications, the IEEE Transactions on Information Theory, and the IEEE Transactions on Information Forensics and Security.
\end{IEEEbiography}

\vspace{-0.1in}
\begin{IEEEbiography}[{\includegraphics[width=1in,height=1.25in,clip,keepaspectratio]{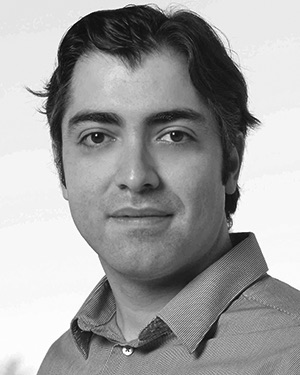}}]{Amin Arbabian} (S’06–M’12–SM’17) received the Ph.D. degree in electrical engineering and computer science from the University of California at Berkeley (UC Berkeley), Berkeley, CA, USA, in 2011. From 2007 and to 2008, he was part of the Initial Engineering Team, Tagarray, Inc., Palo Alto, CA, USA. In 2010, he was with the Qualcomm's Corporate Research and Development Division, San Diego, CA, USA, where he designed circuits for next generation ultralow power wireless transceivers. In 2012, he joined Stanford University, Stanford, CA, USA, as an Assistant Professor of electrical engineering, where he is currently a Frederick E. Terman Fellow with the School of Engineering. His current research interests include high-frequency systems, medical imaging, Internet-of Everything devices including wireless power delivery techniques, and medical implants. 
Dr. Arbabian was a recipient or a co-recipient of the 2015 NSF CAREER Award, the 2014 DARPA Young Faculty Award, the 2013 IEEE International Conference on Ultra-Wideband Best Paper Award, the 2013 Hellman Faculty Scholarship, the 2010 IEEE Jack Kilby Award for Outstanding Student Paper of the International Solid-State Circuits Conference, two-time Second Place Best Student Paper Award at the 2008 and 2011 Radio-Frequency Integrated Circuits (RFIC) Symposium, the 2009 Center for Information Technology Research in the Interest of Society at UC Berkeley Big Ideas Challenge Award and the UC Berkeley Bears Breaking Boundaries Award, and the 2010– 2011 and 2014–2015 Qualcomm Innovation Fellowship. He currently serves on the TPC for the European Solid-State Circuits Conference and the RFIC Symposium.
\end{IEEEbiography}
\vspace{-0.1in}
\begin{IEEEbiography}[{\includegraphics[width=1in,height=1.25in,clip,keepaspectratio]{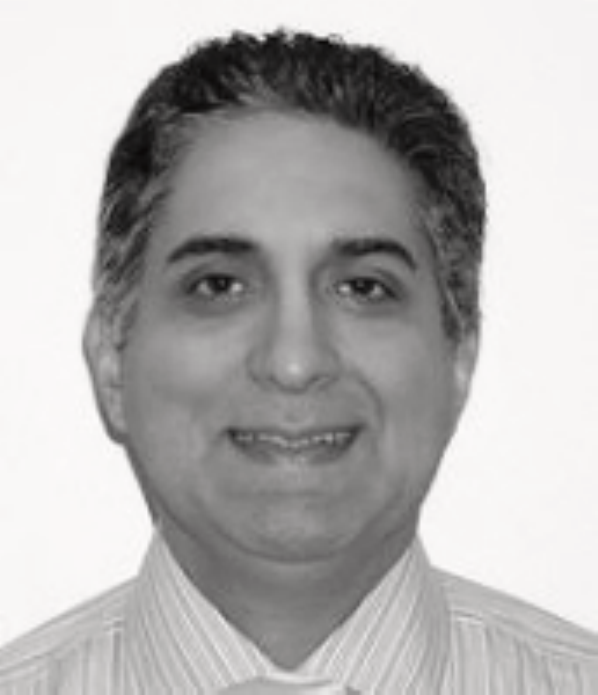}}]{Ali Sadri} is Sr. Director of mmWave Standards and Advanced Technologies at Intel Corporation and the Chairman and CEO of the WiGig Alliance.  Ali has 25 years of engineering, Scientific and academic background in Wireless Communications system.  His Professional work started at IBM in year 1990 and a Visiting Professor at the Duke University through year 2000.  During years 2000-2002 he joined BOPS Inc., a startup company specialized in programmable DSP's as the Sr. Director of the Communications and Advanced Development.  In 2002 Ali joined Intel Corporation's Mobile Wireless Division where he initiated and lead the standardization of the next generation High Throughput WLAN at Intel that became the IEEE 802.11n standards.  Later Dr. Sadri founded and lead the Wireless Gigabit Alliance in 2008 that created the ground breaking WiGig 60 GHz technology.  In June 2013 WiGig Alliance merged with WiFi Alliance to advance and certify the WiGig programs within WiFi alliance framework.  Currently Dr. Sadri is leading the mmWave advanced technology development that includes the next generation WiGig standards and mmWave technology for 5G cellular systems. Ali holds more than 100 Issued and pending patent applications in wired and wireless communications systems.
\end{IEEEbiography}




\end{document}